\documentclass[12pt]{article}
\usepackage[dvips]{color}
\usepackage{epsfig}
\usepackage{amsmath}
\usepackage{graphicx}

\begin{document}
\title{\bf Strong Gravitational Lensing by the R-Charged Non-Extremal Black Hole}
\author{{J. Naji$^{a,}$\thanks{Email: naji.jalil2020@gmail.com}}\\
$^{a,}${\small {\em Physics Department, Ilam University, Ilam, Iran}}\\
{\small {\em P.O. Box 69315-516, Ilam, Iran}}} \maketitle
\begin{abstract}
\noindent In this paper, gravitational lensing scenario due to the R-charged black hole of five dimensional supergravity investigated. We study the effective potential of traveling photons near the R-charged black hole and find some stable orbits for the photons. We also find that the effect of the black hole charges is increasing of the effective potential. We have shown that photons do not cross the horizon of the very large R-charged black hole. By using the numerical study we find that the black hole charges and non-extremality parameter decrease value of the deflection angle.\\\\
{\bf Keywords:} Black hole, Gravitational Lensing.\\\\
{\bf PACs:}
\end{abstract}
\section{Introduction}
Gravitational lensing, predicted by Einstein, is one of the important aspects of general relativity. Gravitational field of a massive object is cause of the bending of light which is large due to strong gravitational field of a black hole, while is small due to weak gravitational lensing. Theory and applications of weak gravitational lensing reviewed by the Ref. \cite{1}. Deflection angle of light due to a strong gravitational field of
the Schwarzschild black hole was studied for the first time by the Ref. \cite{2} and developed by the Refs. \cite{3} and \cite{4}. The strong lensing scenario
by a spherically symmetric black hole \cite{5,6} and a spinning black hole \cite{7,8} has already been studied. Recently, it has been argued that Kerr black hole acting as a gravitational deflector within the geometrical optics \cite{9} and strong gravitational lensing in Kerr space-time including polarization has been studied \cite{10}. Strong gravitational lensing in a squashed Kaluza-Klein black hole in the G\"{o}del like universe with global rotation has been studied by the Ref. \cite{11}. Strong gravitational lensing effect around a dilaton black holes in an anti de Sitter space considered by the Ref. \cite{12}, while gravitational lensing effects of a Reissner–Nordstrom–de Sitter black hole has been studied by the Ref. \cite{13}. The gravitational lensing scenario due to Schwarzschild-like black hole surrounded by quintessence has been investigated by the Ref. \cite{13-1} and special cases of Kiselev black hole has been discussed. Deflection angle of strong gravitational lensing by one-charged STU black hole has been calculated by the Ref. \cite{14}. STU black hole is solution of $\mathcal{N}=2$ five dimensional supergravity which is special kind of charged black hole with three scalar field and it is important from AdS/CFT correspondence point of view \cite{15,16,17,18,19,20,21,22,23}. STU model has generally 8-charged black hole and has many attention in recent theoretical studies \cite{24,25,26,27}.  Therefore, it is quite reasonable to consider R-charged black hole which is special kind of STU model with three electrical charges, and investigate strong gravitational lensing scenario. In this paper we would like to consider five dimensional STU model with three electrical charges as R-charged black hole and calculate photon path and bending angle to find effect of the black hole charge (corresponding to the chemical potential) on the gravitational lensing. Already, properties of null geodesics around an electrically charged, asymptotically flat black hole in Eddington-inspired Born-Infeld gravityhas been studied and found  that the radius of the unstable circular orbit decreases with the coupling constant for the fixed electrical charge \cite{28}. Hence, variation of deflection and position angle, and the magnification due to the light bending in strong gravitational field due to the black hole charge is interesting problem which considered in this paper.\\
This paper is organized as follows. In section 2 we study null geodesics and discuss about the effective potential to find effect of the black hole charges on the photon path which studied in the section 3. In section 4 we calculate deflection angle affected by the black hole charge. Finally in section 5 we give conclusion.

\section{Null geodesics}
\begin{figure}[h!]
 \begin{center}$
 \begin{array}{cccc}
\includegraphics[width=50 mm]{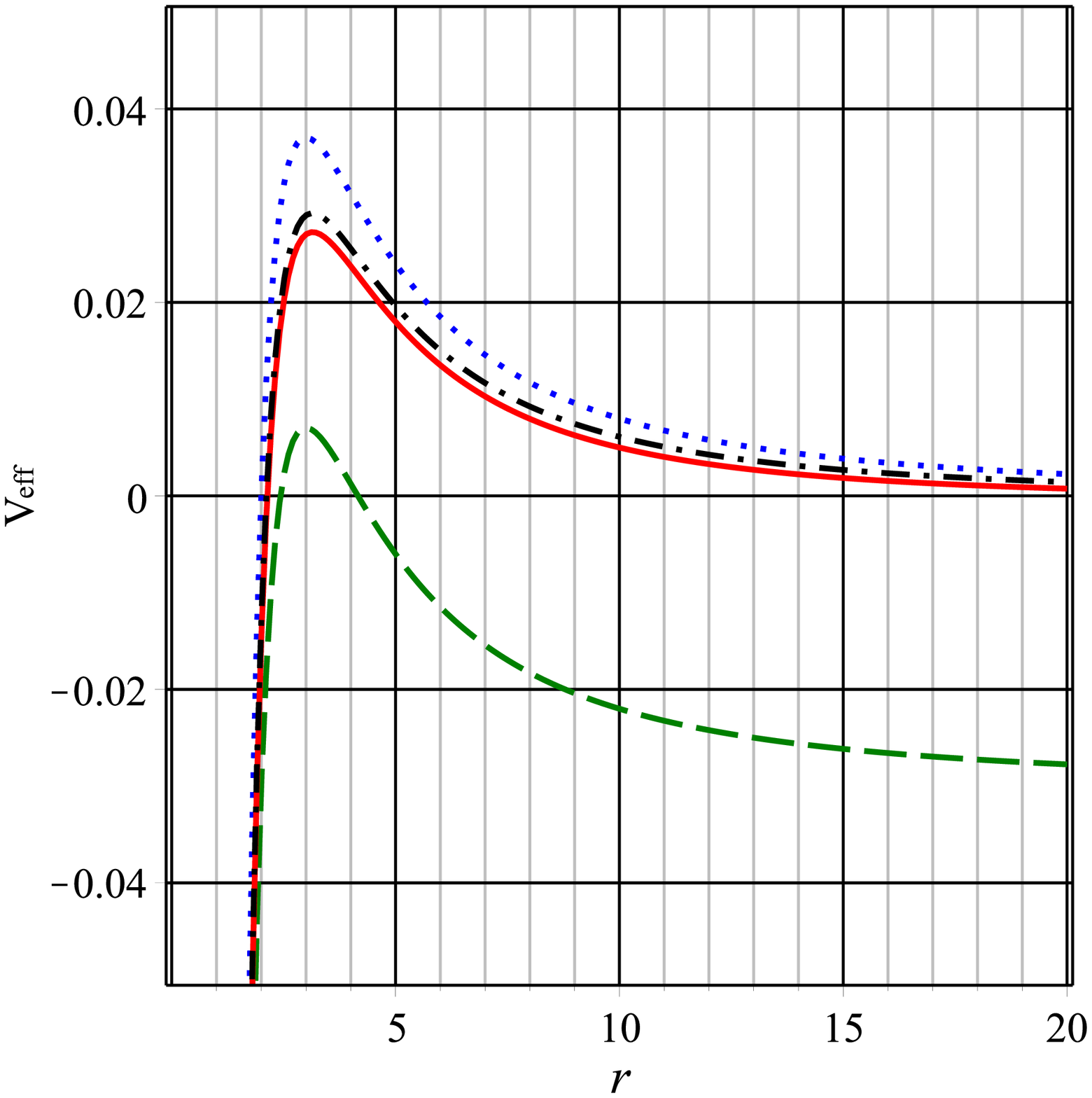}\includegraphics[width=50 mm]{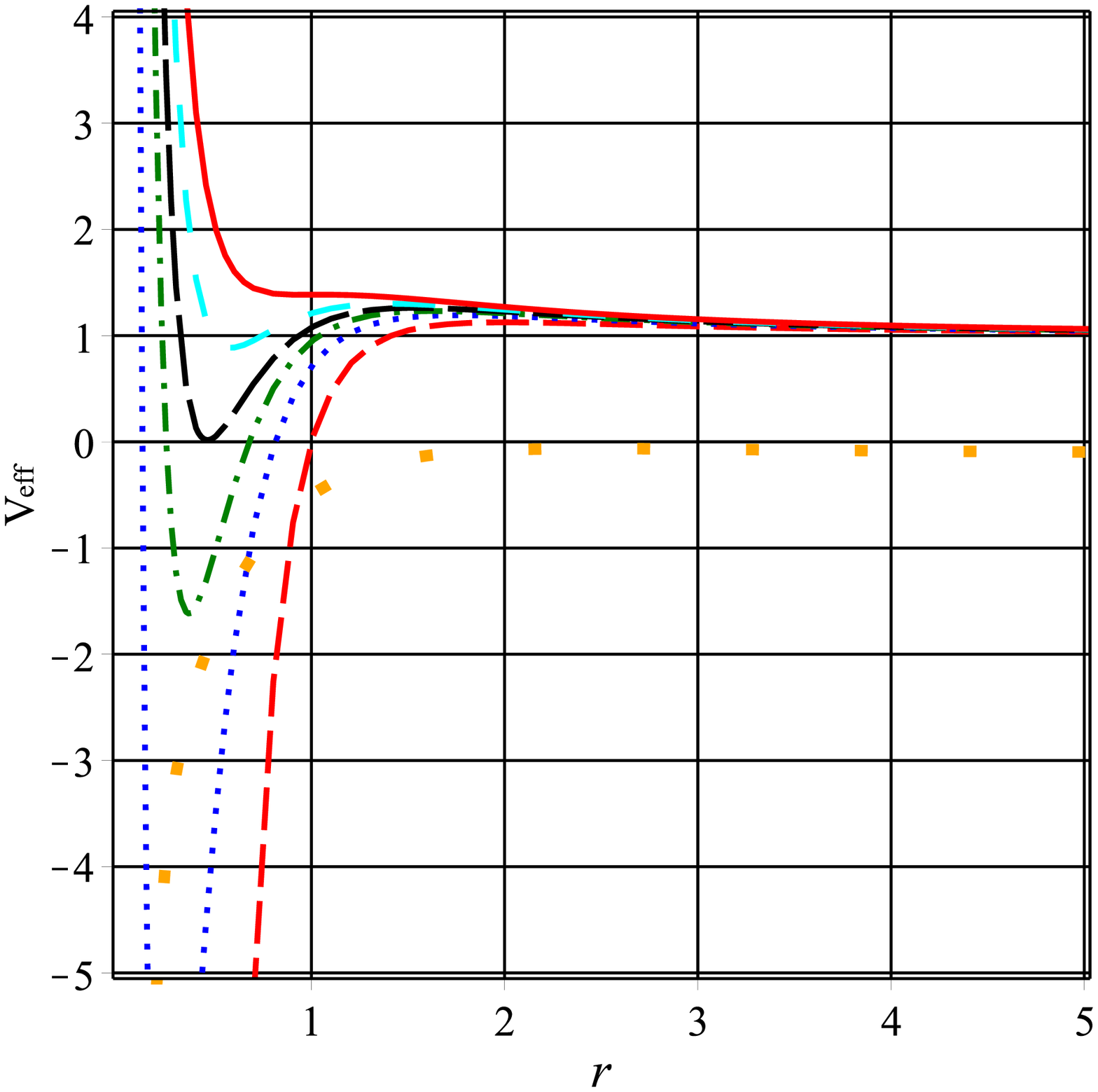}
 \end{array}$
 \end{center}
\caption{Effective potential in terms of $r$ for $L=1$. Left: $M=1$, Schwarzschild (blue dot) Schwarzschild-de-Sitter (green dash) Schwarzschild-quintessence (red solid), Kiselev with $\omega_{q}=-0.6$ (black dash dot) for $\sigma=0.03$. Right: STU with $R^2=1$, $\mu=2$ for $q_{1}=q_{2}=q_{3}=0$ (red dash), $q_{1}=q_{2}=q_{3}=0.3$ (blue dot), $q_{1}=0.5$, $q_{2}=q_{3}=0.4$ (green dash dot), $q_{1}=0.517$, $q_{2}=0.506$, $q_{3}=0.517$ (black long dash), $q_{1}=q_{2}=q_{3}=0.6$ (cyan space dash), $q_{1}=q_{2}=q_{3}=0.722$ (red solid); $q_{1}=q_{2}=q_{3}=1$, $R^{2}=-8$ (orange space dot).}
 \label{fig1}
\end{figure}

The geometry of three-charged non-extremal black hole solution in ${\mathcal{N}}=2$ gauged supergravity known as R-charged black hole (STU black hole) is given by \cite{30},
\begin{equation}\label{1}
ds^{2}=-\frac{f}{{\mathcal{H}}^{\frac{2}{3}}}dt^{2}
+{\mathcal{H}}^{\frac{1}{3}}(\frac{dr^{2}}{f}+r^{2}\left(d\rho^{2}+\sin^{2}\rho d\theta^{2}+\sin^{2}\rho\sin^{2}\theta d\phi^{2})\right),
\end{equation}
where,
\begin{eqnarray}\label{2}
f&=&1-\frac{\mu}{r^{2}}+\frac{r^{2}}{R^{2}}{\mathcal{H}},\nonumber\\
{\mathcal{H}}&=&H_{1}H_{2}H_{3},\nonumber\\
H_{i}&=&1+\frac{q_{i}}{r^{2}},
\end{eqnarray}
with $i=1, 2, 3$ corresponding to three real scalar fields, $X^{i}={\mathcal{H}}^{\frac{1}{3}}/H_{i}$, and $R$ is the constant AdS radius which is equal to the inverse of coupling
constant $g$, ($R=1/g$). The
black hole event horizon denoted by $r=r_{+}$ which is obtained by the following equation,
\begin{equation}\label{3}
1-\frac{\mu}{r^{2}}+\frac{r^{2}}{R^{2}}{\mathcal{H}}=0,
\end{equation}
where $\mu$ is called the non-extremality parameter related to the electrical charges ($q_{i}$) of the black hole.\\
Real solutions of the equation (\ref{3}) are given by,
\begin{equation}\label{4}
r_{\pm}=1\pm\frac{\sqrt{W^{2}-2\mathcal{A}W+4(3\mathcal{B}+\mathcal{A}^{2})}}{6W},
\end{equation}
where,
\begin{eqnarray}\label{5}
W^{3}&=&-36\mathcal{A}\mathcal{B}-108\prod_{i=1}^{3}q_{i}-8\mathcal{A}^{3}\nonumber\\
&+&12\sqrt{-12\mathcal{B}^{3}-3\mathcal{A}^{2}\mathcal{B}^{2}+54\mathcal{A}\mathcal{B}\prod_{i=1}^{3}q_{i}+81(\prod_{i=1}^{3}q_{i})^{2}+12\mathcal{A}^{3}\prod_{i=1}^{3} q_{i}},
\end{eqnarray}
with
\begin{eqnarray}\label{6}
\mathcal{A} &=& q_{1}+q_{2}+q_{3}+R^{2},\nonumber\\
\mathcal{B} &=& \mu R^{2}-q_{1}q_{2}-q_{2}q_{3}-q_{1}q_{3}.
\end{eqnarray}
Comparison with the expression presented by the Ref. \cite{22}, we add the number one in right hand side of the equation (\ref{5}) under assumption of $\Xi\equiv\frac{\sqrt{W^{2}-2\mathcal{A}W+4(3\mathcal{B}+\mathcal{A}^{2})}}{6W}\gg1$, which means large black hole, to have regular solutions. Consideration of large black hole charge yields to validity of our assumption. In that case the $r_{+}$ denotes outer horizon interpreted as black hole event horizon, while the $r_{-}$ denotes inner
horizon.\\
The Lagrangian for traveling of a photon restricted in the equatorial plan ($\theta=\frac{\pi}{2}$) in STU background is given by,
\begin{equation}\label{7}
\mathcal{L}=-\frac{f}{{\mathcal{H}}^{2/3}}{\dot{t}}^{2}+\frac{{\mathcal{H}}^{1/3}}{f}{\dot{r}}^{2}
+r^{2}{\mathcal{H}}^{1/3}{\dot{\rho}}^{2}+r^{2}{\mathcal{H}}^{1/3}\sin^{2}{\rho}{\dot{\phi}}^{2}.
\end{equation}
Under assumption of $\rho=\frac{\pi}{2}$, the Lagrangian (\ref{7}) becomes,
\begin{equation}\label{8}
\mathcal{L}=-\frac{f}{{\mathcal{H}}^{2/3}}{\dot{t}}^{2}+\frac{{\mathcal{H}}^{1/3}}{f}{\dot{r}}^{2}
+r^{2}{\mathcal{H}}^{1/3}{\dot{\phi}}^{2}.
\end{equation}
Using the Euler-Lagrange equations for the null geodesics one can obtain,
\begin{equation}\label{9}
\dot{t}=\frac{E {\mathcal{H}}^{2/3}}{f},
\end{equation}
where constant $E$ denotes energy, and
\begin{equation}\label{10}
\dot{\phi}=\frac{L}{r^{2}{\mathcal{H}}^{1/3}},
\end{equation}
where constant $L$ denotes angular momentum. And finally we have,
\begin{equation}\label{11}
\dot{r}=L\sqrt{\frac{{\mathcal{H}}^{1/3}}{b^{2}}-\frac{1}{r^{2}}\frac{f}{{\mathcal{H}}^{2/3}}},
\end{equation}
where $b=\frac{L}{E}$ is the impact parameter for photons of finite rest mass (which will be calculated using photon path in the next section), and we used the following null condition of the velocity,
\begin{equation}\label{12}
g_{\mu\nu}\dot{x}^{\mu}\dot{x}^{\nu}=0.
\end{equation}
Know we can write the effective potential for traveling photons in STU background. The geodesics of this motion is a force-free and without acceleration. The gravitational force of black hole encode in the effective potential which is given by,
\begin{equation}\label{13}
V_{eff}=\frac{L^{2}}{r^{2}}\frac{1-\frac{\mu}{r^{2}}
+\frac{r^{2}}{R^{2}}{(1+\frac{q_{1}}{r^{2}})(1+\frac{q_{2}}{r^{2}})(1+\frac{q_{3}}{r^{2}})}}{((1+\frac{q_{1}}{r^{2}})(1+\frac{q_{2}}{r^{2}})(1+\frac{q_{3}}{r^{2}}))^{2/3}}.
\end{equation}
In what follow we reproduce some effective potential obtained from other black holes to compare our results with them.\\
In the case of zero charge ($q_{1}=q_{2}=q_{3}=0$), $R\rightarrow\infty$ and $\mu=2Mr$, the effective potential (\ref{13}) reduced to the effective potential of the Schwarzschild black hole given by,
\begin{equation}\label{14}
V_{eff}^{S}=\frac{L^{2}}{r^{2}}(1-\frac{2M}{r}).
\end{equation}
In the case of zero charge ($q_{1}=q_{2}=q_{3}=0$), $R\rightarrow-\frac{1}{\sigma}$ and $\mu=2Mr$, the effective potential (\ref{13}) reduced to the effective potential of the Schwarzschild-de-Sitter black hole given by \cite{31,32,33},
\begin{equation}\label{15}
V_{eff}^{SdS}=\frac{L^{2}}{r^{2}}(1-\frac{2M}{r}-\sigma r^{2}),
\end{equation}
where $\sigma$ is normalization factor.\\
In the case of zero charge ($q_{1}=q_{2}=q_{3}=0$), $R\rightarrow-\frac{r}{\sigma}$ and $\mu=2Mr$, the effective potential (\ref{13}) reduced to the effective potential of the Schwarzschild-like black hole surrounded by quintessence with parameter $\sigma$, which is given by \cite{13-1},
\begin{equation}\label{16}
V_{eff}^{Sq}=\frac{L^{2}}{r^{2}}(1-\frac{2M}{r}-\sigma r).
\end{equation}
In the more general case of zero charge ($q_{1}=q_{2}=q_{3}=0$), $R\rightarrow-\frac{r^{3\omega_{q}+3}}{\sigma}$ and $\mu=2Mr$, the effective potential (\ref{13}) reduced to the effective potential of the Kiselev space-time given by \cite{34},
\begin{equation}\label{17}
V_{eff}^{k}=\frac{L^{2}}{r^{2}}(1-\frac{2M}{r}-\frac{\sigma}{r^{1+3\omega_{q}}}),
\end{equation}
where $\omega_{q}$ is the equation of state parameter for the quintessence scalar field. The Kiselev space-time is the geometry of a static spherically symmetric black hole surrounded by the quintessence.\\
Now, we can effect of the black hole charge on the effective potential and compare results with previous works. In the plots of the Fig. \ref{fig1} we draw effective potential for the traveling photon near R-charged black hole (right) and some important uncharged black holes (left). We can see that effect of the black hole charge and non-extremality parameter is increasing of the effective potential. In the cases of uncharged black holes one can find that there are some situations where photons do not cross the horizon while for the some situations cross the horizon and it is depend on value of $\sigma$ or equivalently $R$. The same situation happen in the case of R-charge black hole. For the sufficient large black hole charges, photons do not cross the horizon. In all cases of uncharged black hole (left panel) there is no stable orbit for the photons, while in the case of R-charged black hole there is some minima corresponding to the stable orbit of photons. In the next section we try to obtain radius of such circular orbit.\\
There is also a critical values for the electrical charges where the following conditions satisfied,
\begin{equation}\label{18}
\frac{dV_{eff}}{dr}=0
\end{equation}
and
\begin{equation}\label{19}
\frac{d^{2}V_{eff}}{dr^{2}}=0,
\end{equation}
which represented by solid red line of the right panel of Fig. \ref{fig1}.

\section{Path equation}
In order to obtain the radius of circular orbit of photons, we use the condition (\ref{18}) and find the following equation,
\begin{eqnarray}\label{20}
&-&(3R+q_{1}+q_{2}+q_{3})r^{10}+(-R(q_{1}+q_{2}+q_{3}-6\mu)-q_{1}^{2}+q_{2}^{2}+q_{3}^{2}-4(q_{1}q_{2}+q_{2}q_{3}+q_{1}q_{3}))r^{8}\nonumber\\
&+&(R(q_{1}q_{2}+q_{2}q_{3}+q_{1}q_{3}+4\mu(q_{1}+q_{2}+q_{3})))r^{6}\nonumber\\
&-&(12(q_{1}q_{2}q_{3})+3(q_{1}q_{2}(q_{1}+q_{2})+q_{1}q_{3}(q_{1}+q_{3})+q_{2}q_{3}(q_{2}+q_{3})))r^{6}\nonumber\\
&+&(3Rq_{1}q_{2}q_{3}+2R\mu(q_{1}q_{2}+q_{2}q_{3}+q_{1}q_{3})-2(q_{1}^{2}q_{2}^{2}+q_{2}^{2}q_{3}^{2}+q_{1}^{2}q_{3}^{2})
-8q_{1}q_{2}q_{3}(q_{1}+q_{2}+q_{3}))r^{4}\nonumber\\
&-&(5q_{1}q_{2}q_{3}(q_{1}q_{2}+q_{2}q_{3}+q_{1}q_{3}))r^{2}-3q_{1}^{2}q_{2}^{2}q_{3}^{2}=0.
\end{eqnarray}
Two real roots of the above equation gives $r_{c\pm}$. In the case of uncharged black hole with the large non-extremality parameter one can find,
\begin{equation}\label{21}
r_{c\pm}(q_{i}=0)=1\pm\sqrt{2\mu},
\end{equation}
hence radius of the stable circular orbit for photons is $r_{ps}(q=0)=\sqrt{2\mu}$ which interpreted as photon sphere.\\
In the case of one-charged black hole ($q_{1}\neq0$, and $q_{2}=q_{3}=0$) with the large $\mu$, one can obtain,
\begin{equation}\label{22}
r_{c\pm}(q_{1}=q)=1\pm\frac{R(6\mu-q)-q^{2}+\sqrt{R^{2}(q^{2}+36q\mu+36\mu^{2})+q^{2}(q^{2}+2Rq+4R\mu)}}{\sqrt{2(3R+q)}},
\end{equation}
therefore radius of photon sphere can be expressed as,
\begin{equation}\label{23}
r_{ps}(q)=\frac{R(6\mu-q)-q^{2}+\sqrt{R^{2}(q^{2}+36q\mu+36\mu^{2})+q^{2}(q^{2}+2Rq+4R\mu)}}{\sqrt{2(3R+q)}}.
\end{equation}
As we can see from right plot of the Fig. \ref{fig1}, for the sufficient large black hole charges there is no minima or maxima, hence there is no stable or unstable circular orbit. However we can have situation with three non-zero charges with two real roots for the equation (\ref{20}). For example with the typical values of $L=R=1$, $\mu=100$, and $q_{1}=q_{2}=q_{3}=0.1$ one can obtain $r_{-}\approx0.003$ and $r_{+}\approx3.060$ while $r_{c-}\approx0.0267$ and $r_{c+}\approx13.4863$. Hence, we can write $r_{-}<r_{c-}<r_{+}<r_{c+}$, and the region of interest is between the two horizons, therefore radius of photon sphere denoted by $r_{c-}$.\\
Now, we are ready to obtain path equation of photons. By using the change of variable $r=\frac{1}{u}$ and chain rule one can obtain,
\begin{equation}\label{24}
(\frac{du}{d\phi})^{2}=\frac{u^{4}\dot{r}^{2}}{\dot{\phi}^{2}}.
\end{equation}
Then, using the equations (\ref{10}) and (\ref{11}) one can obtain path equation for photon as follow,
\begin{equation}\label{25}
(\frac{du}{d\phi})^{2}=(1+q_{1}u^{2})(1+q_{2}u^{2})(1+q_{3}u^{2})(\frac{1}{b^{2}}-\frac{1}{R^{2}})-u^{2}+\mu u^{4}.
\end{equation}
In order to obtain asymptotic behavior ($r\rightarrow\infty$) we set $u\rightarrow0$ and find,
\begin{equation}\label{25-1}
u=\frac{1}{r}=\phi\sqrt{\frac{1}{b^{2}}-\frac{1}{R^{2}}}+Const.,
\end{equation}
which yields to the results of uncharged case at $R\rightarrow\infty$ \cite{13-1}.
In the case of $\frac{du}{d\phi}=0$ one can obtain impact parameter as,
\begin{equation}\label{26}
b=\sqrt{\frac{\mathcal{H}(u)R^{2}}{(u^{2}-\mu u^{4})R^{2}+\mathcal{H}(u)}},
\end{equation}
where $\mathcal{H}(u)=(1+q_{1}u^{2})(1+q_{2}u^{2})(1+q_{3}u^{2})$. Now, by using the change of variable $u^{2}\equiv U$, the equation (\ref{26}) or equivalently right hand side of the equation (\ref{25}) rewrite in the following form,
\begin{equation}\label{27}
AU^{3}+BU^{2}+CU+D=0,
\end{equation}
where
\begin{eqnarray}\label{28}
A&=&\frac{(R^{2}-b^{2})q_{1}q_{2}q_{3}}{b^{2}R^{2}},\nonumber\\
B&=&\frac{(R^{2}-b^{2})(q_{1}q_{2}+q_{1}q_{3}+q_{2}q_{3})}{b^{2}R^{2}}+\mu,\nonumber\\
C&=&\frac{(R^{2}-b^{2})(q_{1}+q_{2}+q_{3})}{b^{2}R^{2}}-1,\nonumber\\
D&=&\frac{R^{2}-b^{2}}{b^{2}R^{2}}.
\end{eqnarray}
In the case of $b^{2}>R^{2}$, appropriate electrical charges, and large $\mu$ one can obtain three real roots (including at least one positive root) of the equation (\ref{27}) denoted by $U_{1}$, $U_{2}$ and $U_{3}$, otherwise we get some imaginary roots. For example in the case of $R=1$, $b=2$, $\mu=100$, $q_{1}=0.5$, $q_{2}=3$ and $q_{3}=5$ one can obtain, $U_{1}\approx-0.05986$, $U_{2}\approx0.14694$ and $U_{3}\approx15.15736$. In the general case we can find one negative root denoted by $U_{1}$ and two other positive roots with the condition $U_{1}<U_{2}<U_{3}$. Therefore, photon path equation (\ref{25}) written in the following form,
\begin{equation}\label{29}
\frac{du}{d\phi}=\pm\sqrt{(u^{2}-u_{1}^{2})(u^{2}-u_{2}^{2})(u^{2}-u_{3}^{2})},
\end{equation}
where positive sign ($\Delta\phi>\pi$) shows photon trajectory toward STU black hole, while negative sign shows photon trajectory away of STU black hole. One can rewrite the equation (\ref{26}) in terms of the closest approach $u_{2}=\frac{1}{r_{cl}}$ (where a photon comes from infinity to reach STU black hole at $r_{cl}$ and again come back to infinity), to obtain relation between $r_{cl}$ and impact parameter $b$ from the following equation,
\begin{eqnarray}\label{31}
&+&\left({R}^{2}-{b}^{2} \right) {r_{cl}}^{12}+ \left( -{b}^{2}R+ \left( -2\,{b}^{2}+{R}^{2} \right)  (q_{1}+q_{2}+q_{3})  \right) {r_{cl}}^{10}\nonumber\\
&+&\left( {b}^{2} ( R\mu-{q_{1}}^{2}-{q_{2}}^{2}-{q_{3}}^{2}) + \left( -4 b^{2}+{R}^{2} \right)  (q_{1}q_{2}+q_{1}q_{3}+q_{2}q_{3})  \right) {r_{cl}}^{8}\nonumber\\
&-&\left( 2b^{2} \left( {q_{1}}^{2}q_{2}+{q_{1}}^{2}q_{3}+q_{1}{q_{1}}^{2}+q_{1}{q_{3}}^{2}+{q_{2}}^{2}q_{3}+q_{2}{q_{3}}^{2}+4q_{1}q_{2}q_{3} \right) -{R}^{2}q_{1}q_{2}q_{3} \right) {r_{cl}}^{6}\nonumber\\
&-&{b}^{2} \left( {q_{1}}^{2}{q_{2}}^{2}+{q_{1}}^{2}{q_{3}}^{2}+{q_{2}}^{2}{q_{3}}^{2}+4q_{1}q_{2}q_{3} (q_{1}+q_{2}+q_{3})  \right) {r_{cl}}^{4}\nonumber\\
&-&2q_{1}q_{2}q_{3}{b}^{2} (q_{1}q_{2}+q_{1}q_{3}+q_{2}q_{3}) {r_{cl}}^{2}-{b}^{2}{q_{1}}^{2}{q_{2}}^{2}{q_{3}}^{2}=0.
\end{eqnarray}
In the case of uncharged black hole ($q_{i}=0$) one can obtain,
\begin{equation}\label{32}
r_{cl}(q_{i}\rightarrow0)=\frac{\left(2b(R^2-b^2)(bR+\sqrt{b^{2}R^{2}+4b^{2}\mu R-4\mu R^{3}})\right)^{\frac{1}{2}}}{2(R^{2}-b^{2})}.
\end{equation}
We can see that (for the appropriate value of $R$) $r_{cl}$ is increasing function of $b$, and the same situation happen for the charged black hole. Effect of the black hole charges are increasing of $r_{cl}$ while non-extremality parameter decreases value of the closest approach.\\
Now, we can calculate deflection angle in the next section.
\section{Deflection angle}
\begin{figure}[h!]
 \begin{center}$
 \begin{array}{cccc}
\includegraphics[width=50 mm]{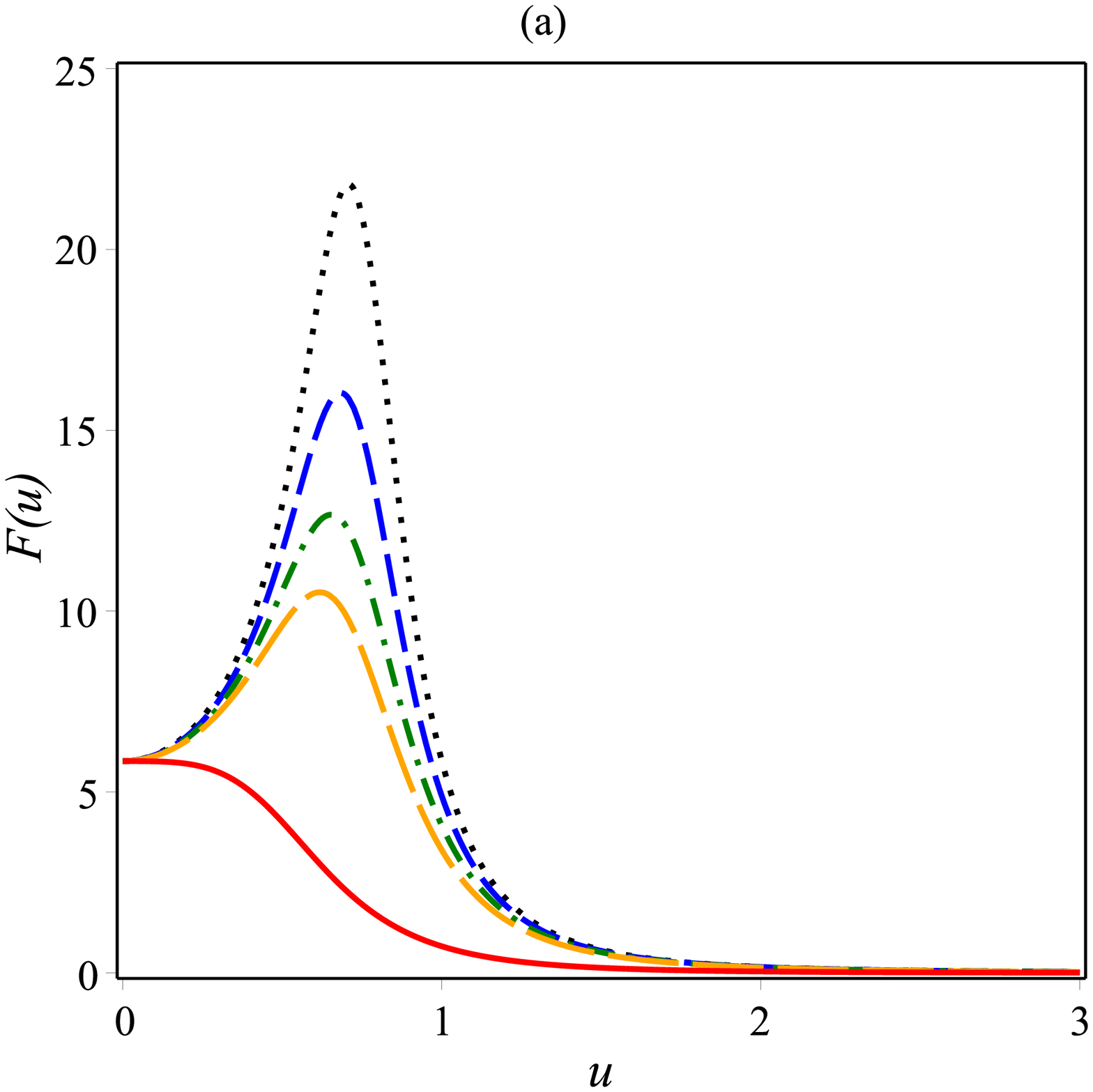}\includegraphics[width=50 mm]{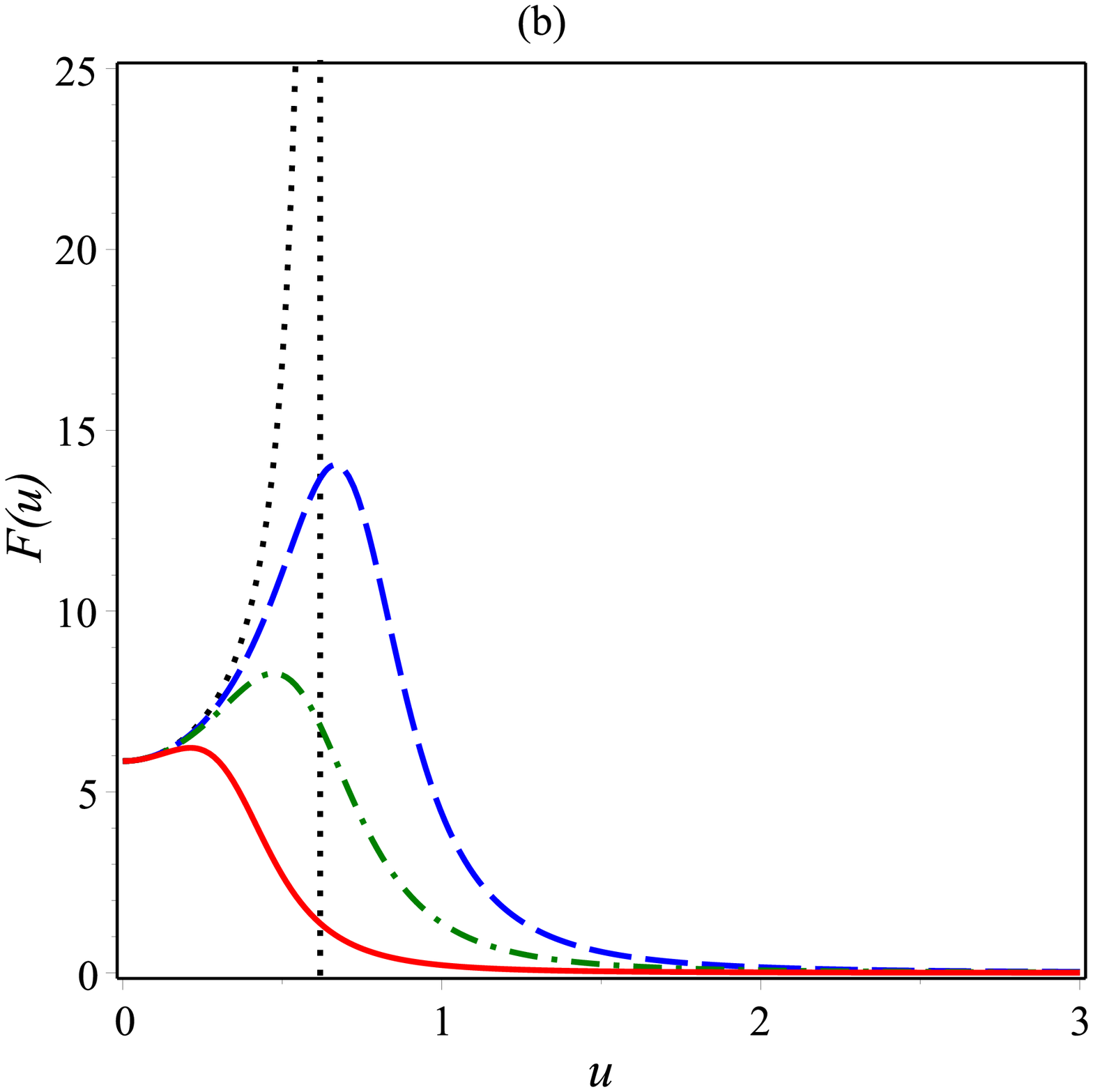}\\
\includegraphics[width=50 mm]{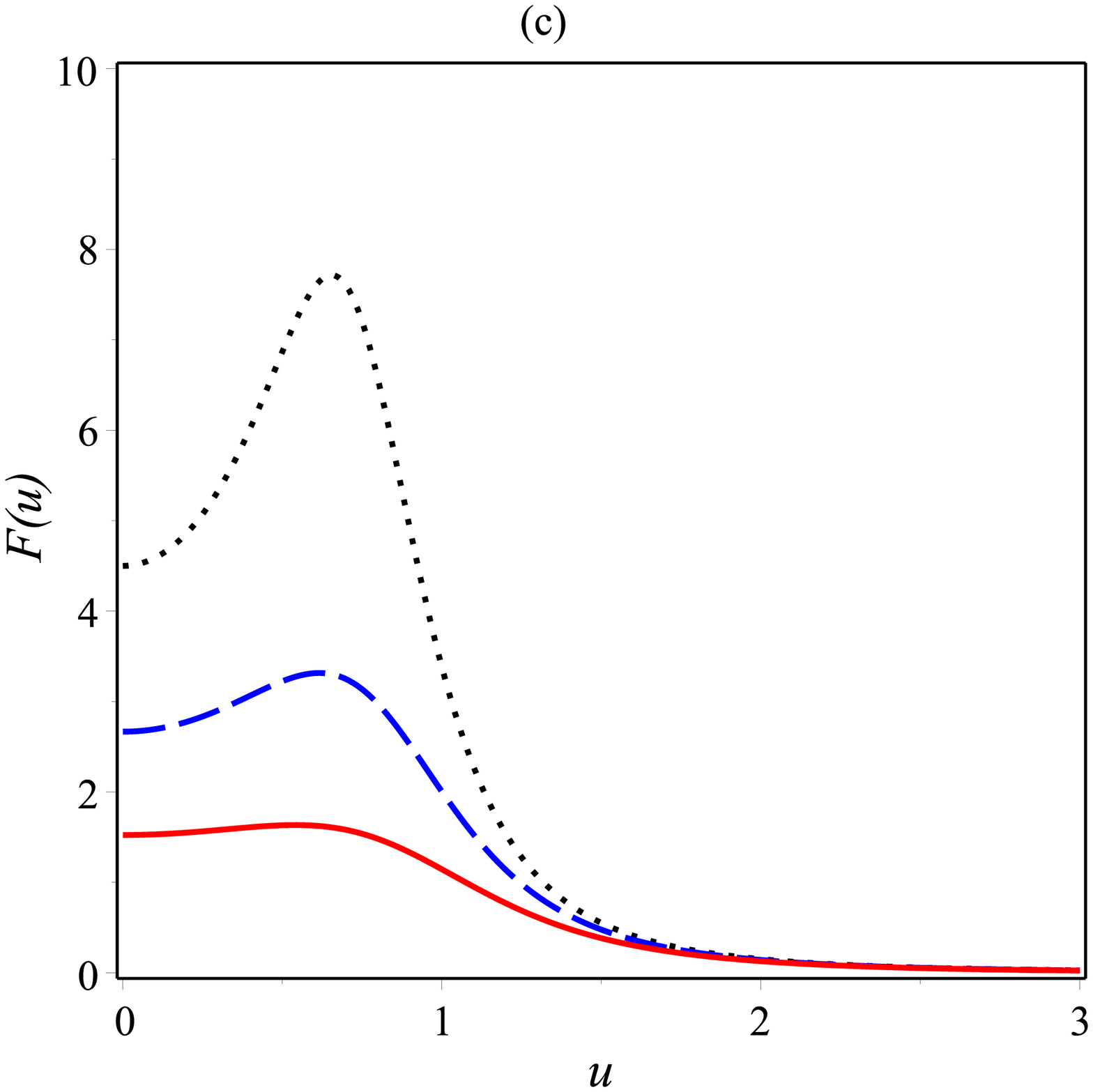}\includegraphics[width=50 mm]{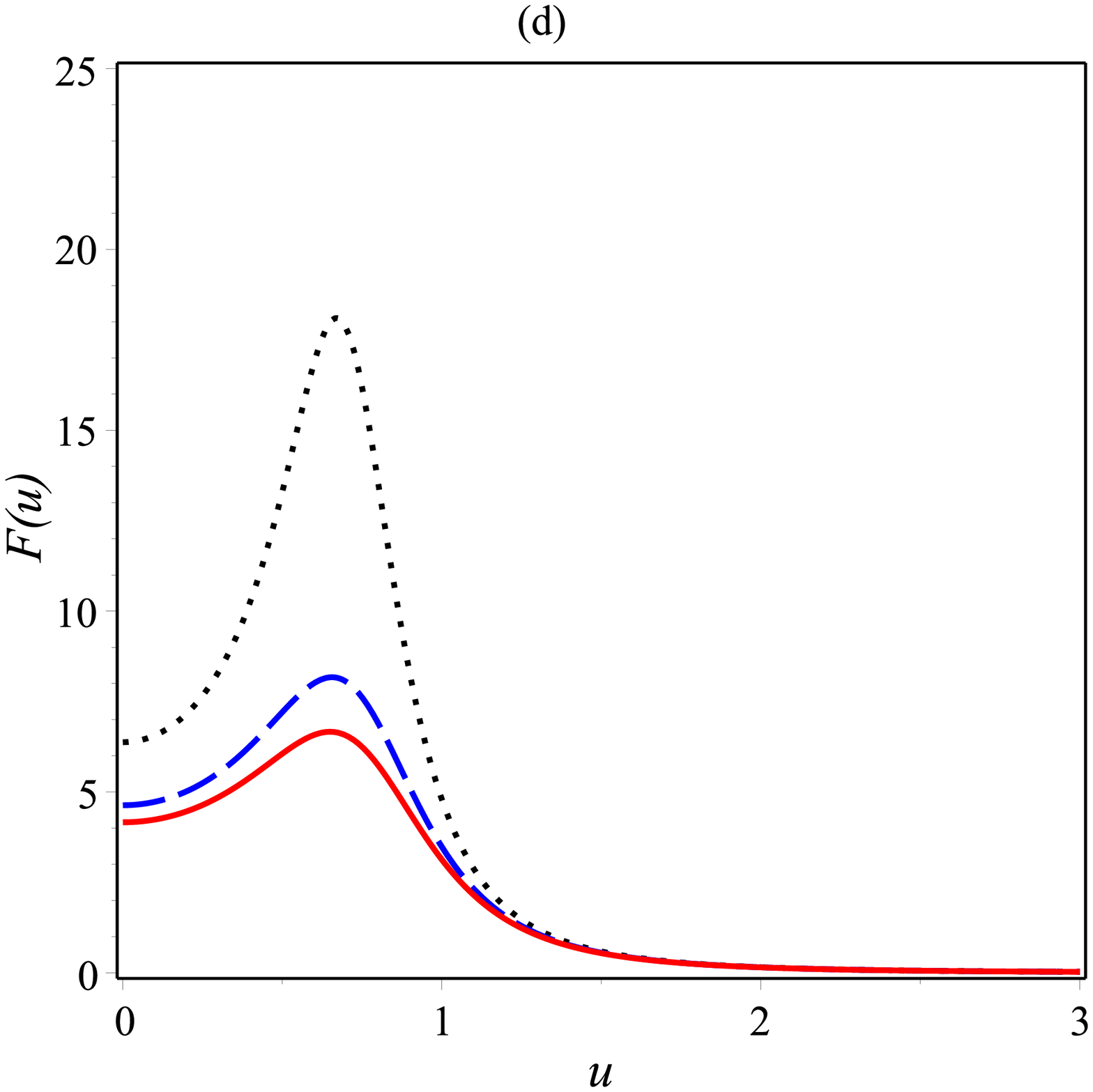}
 \end{array}$
 \end{center}
\caption{$F(u)$ in terms of $u$ with different values of parameters. (a) $b=1.3$, $\mu=1$, $R=2$; $q_{1}=q_{2}=q_{3}=0$ (black dot), $q_{1}=0.2$ and $q_{2}=q_{3}=0$ (blue dash), $q_{1}=q_{2}=0.2$ and $q_{3}=0$ (green dash dot), $q_{1}=q_{2}=q_{3}=0.2$ (orange long dash), $q_{1}=q_{2}=q_{3}=1$ (red solid). (b) $q_{1}=q_{2}=q_{3}=0$, $R=2$, $b=1.3$; $\mu=0$ (black dot), $\mu=1$ (blue dash), $\mu=2$ (green dash dot), $\mu=10$ (red solid). (c) $q_{1}=q_{2}=q_{3}=0.1$, $R=2$, $\mu=1$; $b=1.2$ (black dot), $b=1$ (blue dash), $b=0.8$ (red solid). (d) $q_{1}=q_{2}=q_{3}=0.1$, $b=1.3$, $\mu=1$; $R=1.9$ (black dot), $R=2.5$ (blue dash), $R=3$ (red solid).}
 \label{fig2}
\end{figure}

A photon which comes from infinity near the black hole at $r_{cl}$, deflects and come back to infinity. In that case, deflection angle given by,
\begin{equation}\label{33}
\hat{\alpha}=2\int{d\phi}-\pi=\hat{\alpha}=2\int_{0}^{u_{2}}{\frac{d\phi}{du}du}-\pi,
\end{equation}
where $\frac{d\phi}{du}$ obtained using inverse of the equation (\ref{29}). In that case we can write,
\begin{equation}\label{34}
\hat{\alpha}=2\int_{0}^{\frac{1}{r_{cl}}}{\frac{1}{\sqrt{(u^{2}-u_{1}^{2})(u^{2}-u_{2}^{2})(u^{2}-u_{3}^{2})}}du}-\pi.
\end{equation}
Also, by using the equation (\ref{25}) we can write
\begin{equation}\label{35}
\hat{\theta}=\hat{\alpha}+\pi=\int_{0}^{\frac{1}{r_{cl}}}{F(u)du},
\end{equation}
where
\begin{equation}\label{36}
F(u)\equiv2\left((1+q_{1}u^{2})(1+q_{2}u^{2})(1+q_{3}u^{2})(\frac{1}{b^{2}}-\frac{1}{R^{2}})-u^{2}+\mu u^{4}\right)^{-1}.
\end{equation}
In the Fig. \ref{fig2} we draw $F(u)$ in terms of $u$. Area under curves denote angle $\hat{\theta}$ and we can study the effect of parameters such as the black hole electrical charges, non-extremality parameter and impact parameter on the deflection angle.\\
As the main result we can see from the Fig. \ref{fig2} (a) that effect of the black hole charge is decreasing of the deflection angle as well as the effect of non-extremality parameter (see Fig. \ref{fig2} (b)). Fig. \ref{fig2} (c) shows variation of the deflection angle with impact parameter $b$ and yields to the interesting result; deflection angle increased by $b$ while in the cases of uncharged black hole it is found that the value of $b$ decreases and the deflection angle increases \cite{13-1}. However if we set $q_{i}=0$ and $\mu=0$ recover that results and see that impact parameter decreases value of the deflection angle. Finally in the Fig. \ref{fig2} (d) we can see variation of the deflection angle with $R$ and find that increasing $R$ decreases value of the deflection angle.\\\\

\begin{figure}[h!]
 \begin{center}$
 \begin{array}{cccc}
\includegraphics[width=50 mm]{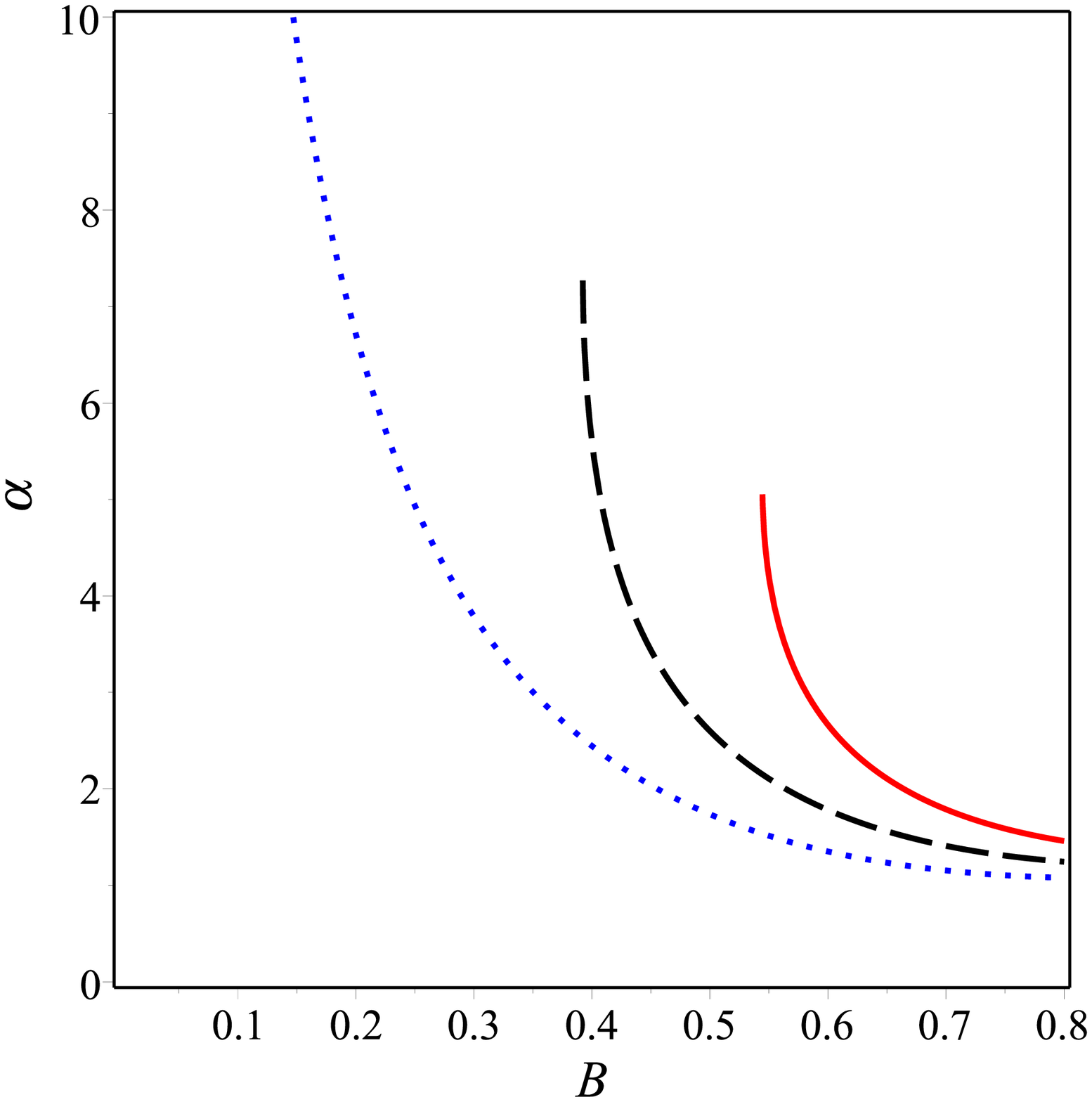}\includegraphics[width=50 mm]{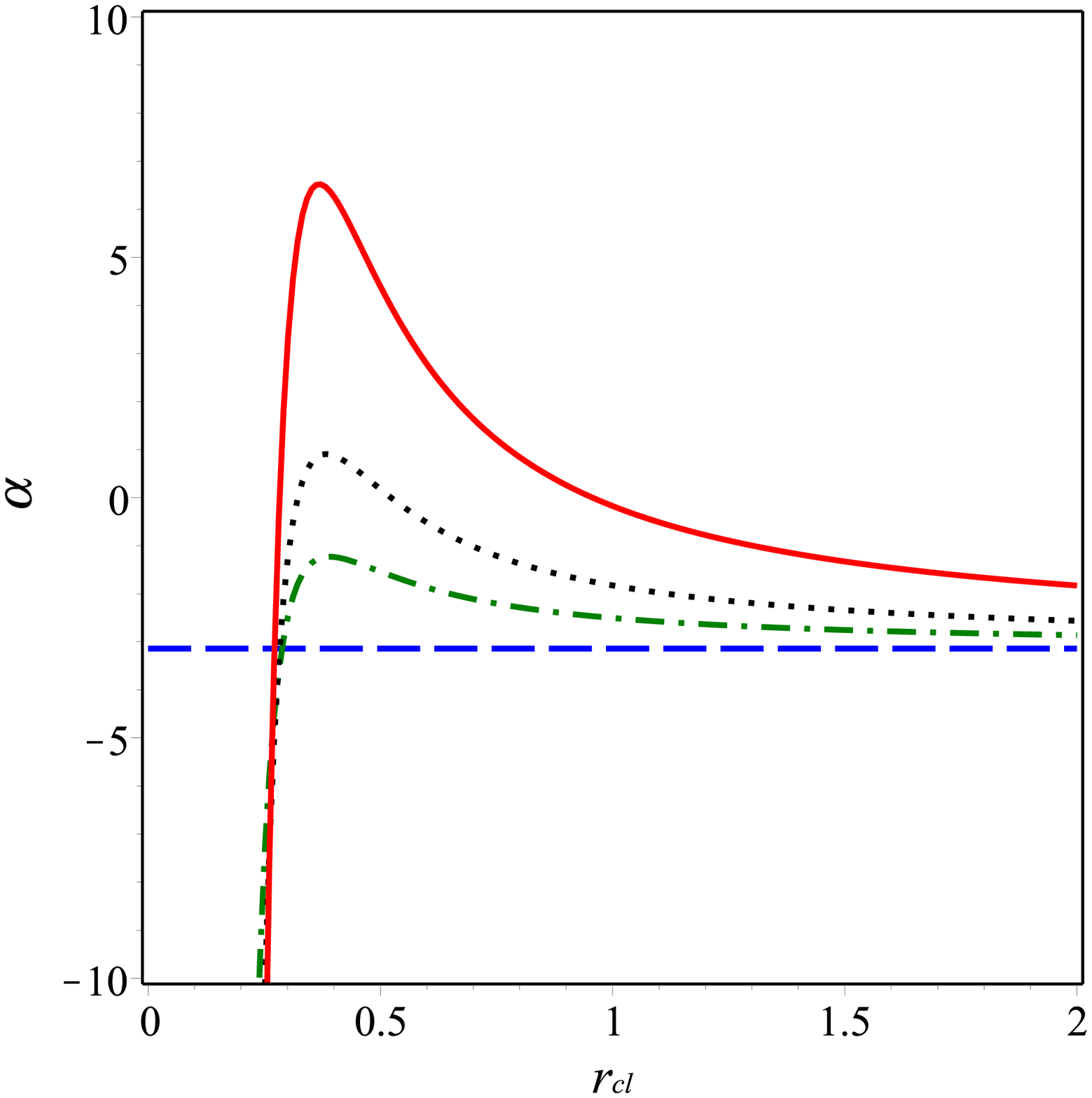}
 \end{array}$
 \end{center}
\caption{Approximate curves of $\hat{\alpha}$ for $R=2$. Left: in terms of $b$ for uncharged black hole with $\mu=0$ (blue dot), $\mu=0.02$ (black dash), $\mu=0.04$ (red solid). Right: in terms of $r_{cl}$ for $q_{1}=q_{2}=q_{3}=\mu=0.1$ with $b=0$ (blue dash), $b=0.5$ (green dash dot), $b=0.7$ (black dot), $b=1$ (red solid).}
 \label{fig3}
\end{figure}

In the case of large $r_{cl}$ and $R$, small $b$, and also infinitesimal charges and non-extremality parameter one can use Taylor expansion and obtain analytical expression for the bending angle $\hat{\alpha}$ as follow,
\begin{eqnarray}\label{37}
\hat{\alpha}\approx-\pi&-&\frac{2}{7}b^{2}\left(1-\frac{{b}^{2}}{{R}^{2}} \right) \frac{q_{1}q_{2}q_{3}}{{r_{cl}}^{7}}\nonumber\\
&-&\frac{2}{5}\frac{b^{2}}{{r_{cl}}^{5}}\left[ (q_{1}q_{2}+q_{1}q_{3}+q_{2}q_{3})\left(1-\frac{{b}^{2}}{{R}^{2}} \right)-\mu \right] \nonumber\\
&+&\frac{2}{3}\frac{b^{2}}{{r_{cl}}^{3}} \left[ 1-(q_{1}+q_{2}+q_{3})\left(1-\frac{{b}^{2}}{{R}^{2}} \right) \right]\nonumber\\
&+&2\frac{b^{2}}{r_{cl}}\left(1+\frac{{b}^{2}}{{R}^{2}} \right).
\end{eqnarray}
For the case of uncharged black hole we yields to the following expression,
\begin{equation}\label{38}
\hat{\alpha}\approx-\pi+2b^{2}\left((1+\frac{{b}^{2}}{{R}^{2}})\frac{1}{r_{cl}}+\frac{1}{3r_{cl}}-\frac{\mu}{5r_{cl}^{5}}\right),
\end{equation}
where $r_{cl}$ given by the equation (\ref{32}).\\
Now we can draw equation (\ref{37}) and see behavior of the bending angle in terms of impact parameter and also closest approach (see Fig. \ref{fig3}). The left plot shows the case of zero-charge black hole, we can see that deflection angle is decreasing function of $b$ and also there is a critical minimum $b_{c}$ where deflection angle exist. For the cases of $b<b_{c}$ there is no bending angle. Right plot show variation of the bending angle with the closest approach. In the case of $b=0$ we can see constant angle. Validity of our approximation to obtain curves of Fig. \ref{fig3} suggests large $r_{cl}$ ($r_{cl}>1$) where the effects of the small black hole charges and non-extremality parameter are negligible.

\section{Conclusion}
In this work, we considered a special kinds of STU black holes which is R-charged non-extremal black hole in five dimensions and study gravitational lensing. Indeed we investigate the effect of the black hole charges on the bending angle and effective potential of the traveling photon in the strong gravitational field of a black hole. As the black hole horizon and therefore size of the black hole is depend on the non-extremality parameter and black hole charges we assumed large black hole. Already, it is found that there is not a stable circular orbit for the traveling photon near uncharged black hole \cite{13-1}, while we found a minimum in the effective potential which shows presence of a stable circular path, hence obtained closest path of photon.\\
By using the numerical study, the effect of the electrical charges and non-extremality parameter of the black hole obtained on the deflection angle and found that it is decreasing function of the black hole charges. We also found that effect of impact parameter on the deflection angle is opposite in the cases of charged and uncharged black hole. Bending angle for the uncharged black hole is decreasing function of the impact parameter while is increasing function of the impact parameter for the R-charged black hole.\\
By using some approximations we obtained analytical expression for the large closest approach (corresponding to the large black hole) and found that the deflection angle is decreasing function of the impact parameter. Effect of non-extremality parameter of this case opposite of the case of small closest approach is increasing of the bending angle.
For the future works it is interesting to consider other kinds of STU black holes like in four or ten dimensions or including magnetic charges \cite{35, 36, 37, 38, 39}.

\end{document}